\documentclass[%
 reprint,
superscriptaddress,
 amsmath,amssymb,
 aps,
]{revtex4-2}
\usepackage{bm}
\usepackage{graphicx}
\usepackage{amssymb}
\usepackage{amsmath}
\usepackage{eufrak}
\usepackage{color}
\usepackage[utf8]{inputenc}
\usepackage[unicode=true,colorlinks=true,urlcolor=blue,citecolor=blue]{hyperref}

\usepackage{pifont}
\usepackage{ulem}
\usepackage[dvipsnames]{xcolor}

\usepackage{graphicx}
\usepackage{dcolumn}
\usepackage{bm}
\usepackage{hyperref}

\usepackage{orcidlink}

\begin{document}

\preprint{APS/123-QED}

\title{Polarized photoluminescence and g-factor in an ensemble of quantum dots in magnetic fields}


\author{Lyubov Kotova~\orcidlink{0000-0001-8767-9252}}\email{kotova@mail.ioffe.ru}
\affiliation{Ioffe Institute, 194021 St.~Petersburg, Russia}
\affiliation{Faculty of Physics, Lomonosov Moscow State University, 119991 Moscow , Russia}

\author{Timur Shamirzaev~\orcidlink{0000-0002-9914-9707}} 
\affiliation{Rzhanov Institute of Semiconductor Physics, Siberian Branch of the Russian Academy of Sciences, 630090 Novosibirsk, Russia}

\author{Vladimir Kochereshko~\orcidlink{0000-0002-5673-8237}} 
\affiliation{Ioffe Institute, 194021 St.~Petersburg, Russia}

\date{\today}

\begin{abstract}
In this paper, polarized photoluminescence caused by exciton quasi-equilibrium spin orientation on Zeeman sublevels in an ensemble of quantum dots of different sizes is theoretically studied. It is found that: (\textit{i}) the splitting of the photoluminescence bands in a magnetic field in an ensemble of quantum dots is several orders of magnitude larger than the Zeeman splitting of exciton levels in a single dot, (\textit{ii}) the sign of the circular polarization of the photoluminescence changes along the contour of the bands, (\textit{iii}) the change of the sign of the polarization is associated with the change of the sign of the exciton g-factor. A universal formula for the dependence of the exciton g-factor on the quantum dot size is obtained. This dependence is valid for both bulk material and nanostructures of different types and sizes. Comparison of the results obtained with the experimentally measured data has been carried out.

\begin{description}
\item[Keywords]
Polarized photoluminescence; quantum dots; g-factor.
\end{description}
\end{abstract}


\maketitle

\section{Introduction}

One of the key parameters for spin quantum information processing is the spin relaxation time, and its study is of great interest for spintronics~\cite{1,WU201061,PhysRevLett.125.156801,RevModPhys.80.1517}. As shown theoretically~\cite{PhysRevB.61.12639,PhysRevB.64.125316,glazov2018electron} and experimentally~\cite{PhysRevB.96.035302,PhysRevB.104.045305}, 
the spin relaxation time of localized carriers and excitons can reach milliseconds due to the suppression of spin-orbit coupling effects.

An effective method for studying exciton and impurity states in crystals is magnetic field induced polarized photoluminescence (MCPL), which is related to the temperature redistribution of carriers and excitons between Zeeman sublevels in a magnetic field. This method has been successfully used as early as Thomas and Hopfield to study bound excitons~\cite{PhysRevLett.7.316}. 
Using this method, the fine structure of the acceptor impurities was investigated and the concentration profile of acceptors~\cite{Uraltsev} in quantum wells was measured. This method has also shown its high efficiency for studying the spin dynamics of carriers and excitons~\cite{Ivchenko2018}.

The population of states split by magnetic field is determined by their g-factor. The size quantization of charge carriers in nanostructures leads to a change of their g-factors. For the electron, this change is due to the spin-orbit interaction, which was first theoretically shown by L. Roth~\cite{PhysRev.114.90} and confirmed by numerous experimental works~\cite{PhysRevB.75.245302}. The mixing of hole states due to the complex structure of the valence band leads to a high sensitivity of the hole g-factor to the shape of the quantizing potential, as was shown in~\cite{PhysRevB.104.205423,16}. For the electron-hole exciton bound state, as shown in~\cite{PhysRevB.78.085204}, there appears an additional contribution to the g-factor due to the lateral motion of the exciton.

\begin{figure}[h!]
\centering
 \includegraphics[width=0.65\linewidth]{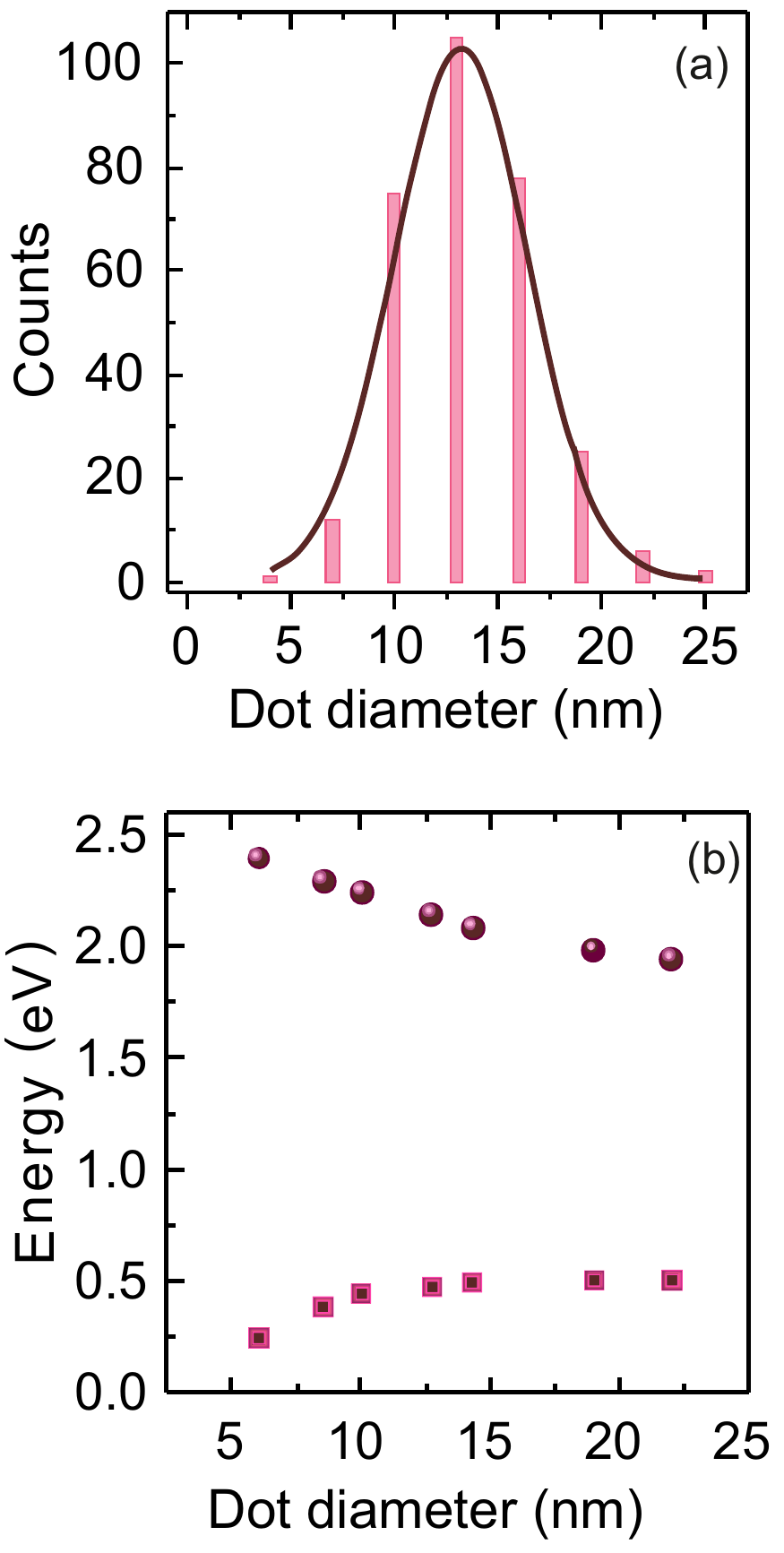}
  \caption{ \textbf{(a)} An example for histogram of the QD-diameter distribution for (In,Al)As/AlAs QDs~\cite{PhysRevB.78.085323}.  The size dispersion fitted by a Gaussian distribution is shown by the line. The aspect ratio is 1:5 for all dots. \textbf{(b)} Energy of electron and heavy hole in (In,Al)As QDs as a function of the QD diameter. The energy is given relatively the top of the AlAs valence band~\cite{PhysRevB.90.125431}.  }
 \label{fig:1}
\end{figure}

    An ensemble of quantum dots, is a system in which the scattering of dot sizes (an example of QD-diameter distribution for (In,Al)As/AlAs QDs~\cite{PhysRevB.78.085323} is shown in Figure~\ref{fig:1}~a leads to a set of objects with different dimensional quantization of electron-holes and excitons (as it shown in Figure~\ref{fig:1}~a for (In,Al)As QDs~\cite{PhysRevB.78.085323}) and, consequently, to a scattering of their g-factors~\cite{PhysRevB.90.125431}.

In this work, the effect of the dispersion of quantum dot sizes in the ensemble on the magnetically induced circular polarization of photoluminescence has been theoretically considered, taking into account the dispersion of the g-factors of holes, electrons, and excitons.

\section{Theory}

\subsection{\label{sec:1}{Polarized luminescence}}

\begin{figure}[h]
\centering
 \includegraphics[width=1\linewidth]{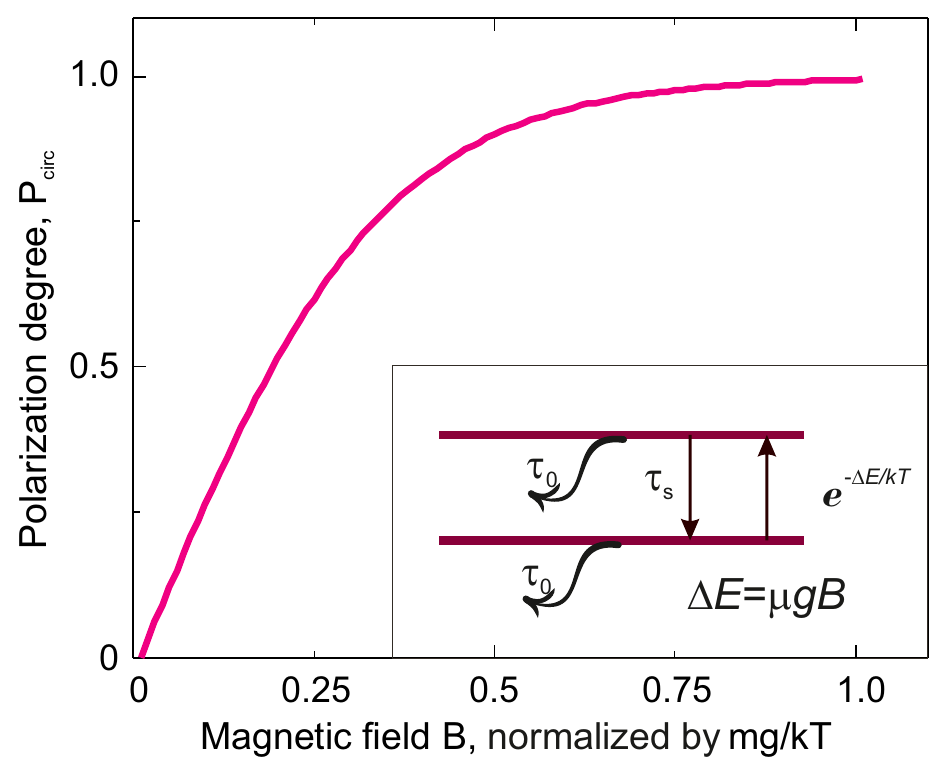}
 \caption{Dependence of the polarization degree of exciton emission on magnetic fields as Eq.~\eqref{eq3} 
 (in units of ${\mu g \over{2kT}}$  , assuming ${\tau_0 \over {\tau_0 + \tau_s}}\equiv1$   , ${\mu g \over{2kT}}\equiv 1$  ). Inset shows the scheme of the transitions.}
 \label{fig:2}
\end{figure}

In a magnetic field, the exciton states split according to the projection of the magnetic moment on the magnetic field direction. Optically active exciton states with zero orbital momentum and total spin momentum equal to $ \mid 1 \rangle  $  are split into two states with angular momentum projection to the magnetic field direction equal to $+1$  or $-1$ . Under non-resonant, unpolarized optical excitation, these states are populated according to the Boltzmann distribution. The population ratio of these levels is determined by the magnitude of the Zeeman splitting $\Delta E$  and the temperature factor  $kT$ (see the inset to Figure~~\ref{fig:2}). In the case of a totally equilibrium distribution this population ration is describe by equation:  

  \begin{equation}
\label{eq1}
	{n_2 \over n_1 }= {e^{-{{\Delta E} \over {kT}}}}.
\end{equation}				 

Here $n_1$  and $n_2$  are the concentrations of excitons at sublevels, $\Delta E (B)=\mu gB$  is the magnitude of Zeeman splitting, $k$  is Boltzmann constant, $g$  is  g-factor,  $B$ is magnetic field, and  $\mu$ is Bohr magneton. 

Emission from these states has right or left circular polarization depending on the sign of the angular momentum projection to the field direction. The intensities of the emission lines are proportional to the population of the levels. The polarization degree of the photoluminescence is equal to: 

  \begin{equation}
\label{eq2}
	P_{\rm circ}=\frac{I_{\sigma_+}-I_{\sigma_-}}{I_{\sigma_+}+I_{\sigma_-}},
\end{equation}				 
where  $I_{\sigma_+}$ and $I_{\sigma_-}$  are radiation intensity in right and left circular polarizations. 

In  the simplest case the splitting of the emission lines is equal to the value of the Zeeman splitting of levels, and the ratio of intensities is determined by the Boltzmann factor. The polarization degree of radiation is equal to: 
  \begin{equation}
\label{eq3}
	P_{\rm circ}={\tau_0 \over {\tau_0 + \tau_s}} th{{\Delta E}\over{kT}}.
\end{equation}				 
Here: $\tau_0$  is lifetime, $\tau_s$  is spin relaxation time, ${\tau_0 \over {\tau_0 + \tau_s}}$ the multiplier   takes into account the fact that at a finite lifetime the full equilibrium does not have time to be established. 

The dependence of the polarization degree on the magnetic field is shown in Figure~\ref{fig:2}.

Experiments on polarized luminescence usually involve a large inhomogeneous ensemble of exciton states. This is especially evident for the ensemble of quantum dots in which there is a large scatter of exciton resonance energies. In this case polarized luminescence acquires some peculiarities.

Let us consider an ensemble of quantum dots. We consider that the exciton luminescence line from a single quantum dot has $\delta$-shape. 

  \begin{equation}
\label{eq4}
L(E,E'){\sim} n(E) \delta (E-E'),
\end{equation}				 
where $E$  is the energy of exciton resonance in a quantum dot, $n(E)$  is population of this dot, $L(E)$  is the emission line shape in a single dot.

In the ensemble of QDs there can be dots of different sizes. The scattering of dots by sizes leads to the scattering of exciton resonance energies $E$ . We assume that the distribution of resonance energies in the ensemble has a Gaussian form: 

  \begin{equation}
\label{eq5}
G(E){\sim} e^{-\left({{E-E_0} \over w }\right)^2},	
\end{equation}				 
here $E_o$  is the most probable exciton energy in the ensemble and $\sigma$ is the dispersion. 

Then the luminescence band shape of the QD ensemble is a convolution of  $\delta$-shaped luminescence lines from each QD and the Gaussian distribution: 

  \begin{equation}
\label{eq6}
I(E')=\int_{-\infty}^{\infty} G (E) L (E,E') \,dE.	
\end{equation}				 

As a result, we get: 

  \begin{equation}
\label{eq7}
I(E') \sim	e^{-\left({{E'-E_0} \over {w}}\right)^2} n(E').
\end{equation}				 

In a magnetic field, the states are split in energy according to the projection of the magnetic moment on the direction of the magnetic field. The energies of these states in the magnetic field will be as follows: 

  \begin{equation}
\label{eq8}
E'\rightarrow E' \pm {1 \over 2} \mu B g_{eff} (E').	
\end{equation}				 

The split levels will be populated according to the Boltzmann distribution. As a result, for the emission line of the QD ensemble in a magnetic field we obtain: 

  \begin{equation}
\label{eq9}
I(E') \sim	e^{-\left({{E'-E_0} \over {w}}\right)^2 \pm {{\Delta E (E')} \over {kT}}}
\end{equation}				 

\begin{widetext}

To simplify the formulas, let us put $E_0=0$ . The position of the emission band maximum is determined from the solution of following equation: 

  \begin{equation}
\label{eq10}
\left[ E' \pm \mu B g_{eff} (E') \right] \left[	1 \pm \mu B {{d g_{eff} (E')}\over dE'} \right] \mp {{w^2} \over {2kT}} \mu B {{d g_{eff} (E')}\over dE'} = 0.
\end{equation}				 

This equation can have several roots. Consequently, the shape of the bands will be different from the Gaussian one and will have several maxima. The bands intersect when $I_{{\sigma_+}}=I_{{\sigma_-}}$ , i.e., when the value of the Zeeman splitting $\Delta E (E')$  in the exponent in Eq.~\eqref{eq9} goes to zero. If the intersection points of the bands fall in the energy region where the intensity of the bands has a "measurable" value, the polarization $P_{\rm circ}(E')$  changes sign at these points. 

\end{widetext}

Let's look at a few simple cases.

I.  If $g_{eff}$  is \underline{independent on energy $E'$} : $g_{eff}=g_0$. Then: 

 \begin{figure}[h]
\centering
 \includegraphics[width=1\linewidth]{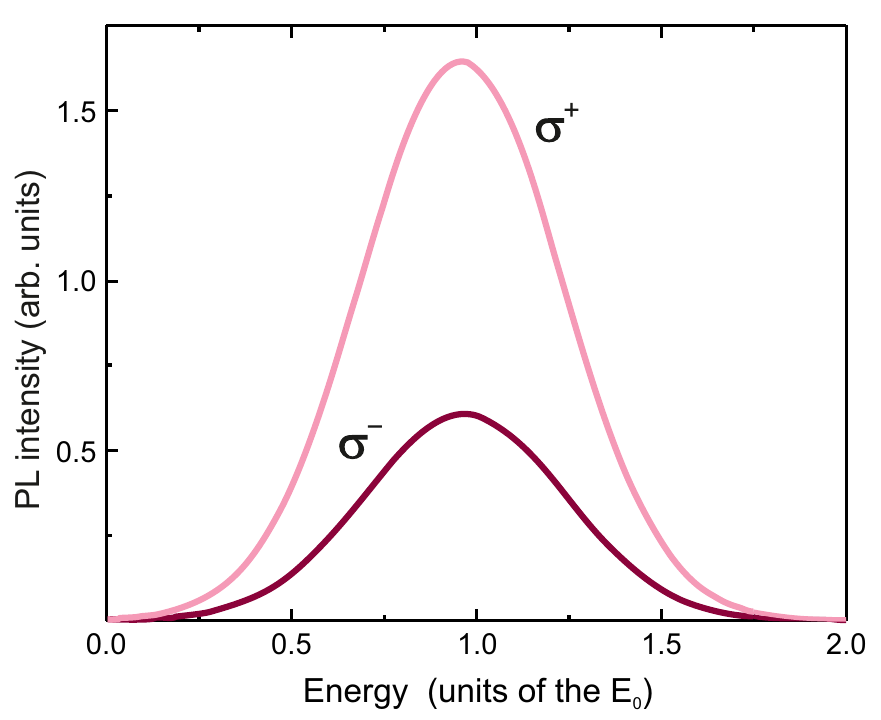}
 \caption{Qualitative view of the emission spectrum of an ensemble of quantum dots in two circular polarizations $\sigma_+$  and $\sigma_-$  as a function of the exciton quantization energy in a dot (in the units of the central energy $E_0$  in Eq.~\eqref{eq7}) under the assumption that the g-factor does not depend on the exciton quantization energy. 
}
 \label{fig:3}
\end{figure}

\begin{enumerate}
\item	The maxima of emission bands in two circular polarizations are at energies $E'_{\pm}={1 \over 2} \mu g_0 B$ .  
\item	The ratio of amplitudes of these bands is equal to  $e^{{{\mu B}\over {kT}} g_{eff}}$ . 

\item	The more intense emission band is below the less intense one in energy. That agrees with the Boltzmann distribution of excitons at sublevels. 

\item The width at half-height of both strips is equal to $2 w \sqrt{ln 2}$ . 

\item The bands intersect at $E'= {w^2 \over {2kT}}$  . Since $w\gg kT$ , the point of intersection is far from the maxima of the bands.

\item  The point of intersection of the band contours is above both maxima in energy. 

\item The degree of polarization does not change sign along the contour of the emission bands. 

\end{enumerate}
The spectral dependence of the enission intensities in two circular polarizations is shown in Figure~\ref{fig:3}. The parameters of this calculation are:  $w=20$,  $E_0=50$,  $\mu g B=0.25$.

II.  Assume that  $g_{eff}$ \underline{ depends linearly on the energy}: $g_{eff} (E')=g_0 +  \Tilde{g}E'$ . Let's estimate $\mu B {{d g_{eff} (E')}\over dE'}$. In the field of 10~T $\mu B {{d g_{eff} (E')}\over dE'} \approx 0.053 \cdot 10 \cdot \Tilde{g}$. It is reasonable to consider that g-factor can vary from -10 to 10 in the energy range of 100~meV. Then $\mu B {{d g_{eff} (E')}\over dE'} \ll 1$  and in Eq.~\eqref{eq8} this 
summand can be neglected compared to unit. 

\begin{figure}[b]
\centering
 \includegraphics[width=1\linewidth]{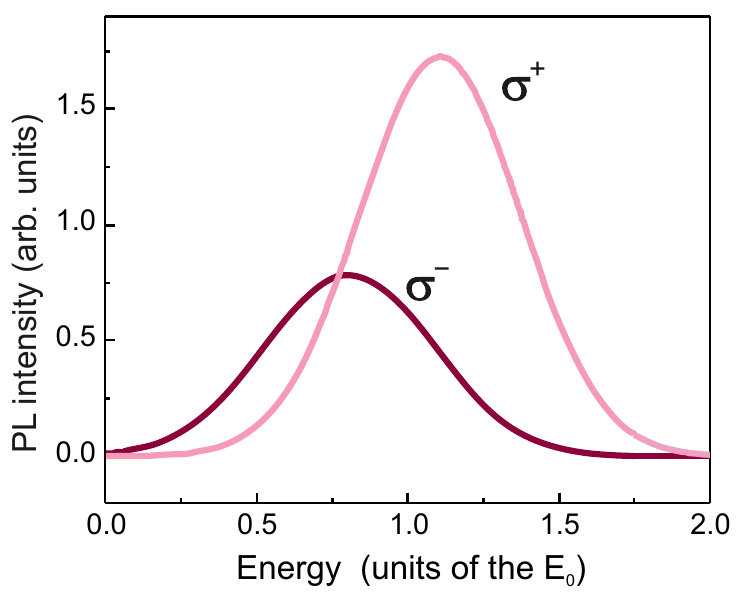}
 \caption{Qualitative view of the emission spectrum of an ensemble of quantum dots in two circular polarizations as a function of the exciton quantization energy in a QD (in the units of the central energy $E_0$ in Eq.~\eqref{eq7}) under the assumption that the $g$-factor depends linearly on the exciton quantization energy $g_{eff} (E')=g_0 +  \Tilde{g}E'$. }
 \label{fig:4}
\end{figure}

\begin{enumerate}
\item The maxima of emission bands in two circular polarizations are at energies
\begin{equation}
\label{eqn11}
E^{\pm}_{max}= {{\mu B \left( \mu B g_0 \Tilde{g} \mp g_0 \pm {w^2 \over{2kT}}\Tilde{g} \right)} \over {(1 \pm \mu B \Tilde{g})}^2}.
\end{equation}

Note that the ratio  ${w \over{kT}}$  can reach values $ \sim 10^2 - 10^3.$ Hence, approximately $E^{\pm}_{max} \approx \pm \mu B {w^2 \over{2kT}} \Tilde{g} $.In this case, the splitting of the emission bands of the ensemble of QDs can be very large compared to the Zeeman splitting of the exciton emission lines in a single QD.

\item Ratio of amplitudes of emission lines  $ \sim e^ {{\left( {\sigma^2 \over{kT}} \mu B \Tilde{g} \right)}}$.

\item Amplitudes of the bands are: 

\begin{equation}
\label{eqn12}
I^{\pm}_{max}= e^{-\{{{(E^{\pm}_{max})^2} \mp {{{\mu B w^2 g (E^{\pm}_{max})}\over{2kT}}}\over w^2}\}}.
\end{equation}

It can be seen that the ratio of band intensities depends on the sign of the g-factor on the band maximum energy. If $g(E_{max})<0$  the more intense emission band is above the less intense one in energy, which is inconsistent with the Boltzmann distribution of excitons on sublevels (Figure~\ref{fig:4}). If $g(E_{max})>0$  the picture is similar to the case of a constant g-factor (Figure~\ref{fig:3}). 

 
\item The widths of these two emission bands $\Delta_\pm$  are different in this case:

  \begin{equation}
\label{eq11}
	\Delta_\pm = 2 w \sqrt{ln 2 \mp {1\over{2kT}}\mu B\Tilde{g}}
\end{equation}				 
The less intense band is slightly wider than the more intense band. The difference in band widths is less than 10\%. 

\item The crossing of the bands and, accordingly, the zeroing of the degree of polarization of radiation occurs at the energy when, either $g_{eff} (E')=g_0 +  \Tilde{g}E'=0$ , or $E'={w^2 \over{2kT}}$ . 

\item It is obvious that $g_{eff}(E')$ can becomes zero only if the g factor $g_{eff}(E)$  changes sign.

\item The degree of polarization along the contour of the emission band changes sign at the energy where $g_{eff}(E')$ changes sign. 

The spectral dependence of the enission intensities in two circular polarizations is shown in Figure~\ref{fig:4}. Parameters of this calculation are:  $w=20$,  $E_0=50$,  $\mu g_0 B=0.8$, $\mu \Tilde{g} B =0.02(meV)^{-1}.$  
\end{enumerate}

III. The \underline{ realistic dependence of a $g$-factor on energy} is discussed in paragraph~\ref{sec:G}.

Let us assume that $g_{eff}$  is determined by the formula: 

  \begin{equation}
\label{eq12}
{g_{eff}	(E')=g_0+ {{aE'}\over{bE'+R}}}.
\end{equation}				 
Here:  $g_0, a  ,   b$ and $R$  are some constants. For the model calculation we use  $a=0.12 (meV)^{-1}$,  $b=0.5 \cdot a$, $R=3 meV $. 

Note that in  Eq.~\eqref{eq12} these quantities are scaled and nothing depends on their absolute values. 

\begin{figure*}[t]
\centering
\includegraphics[width=1\linewidth]{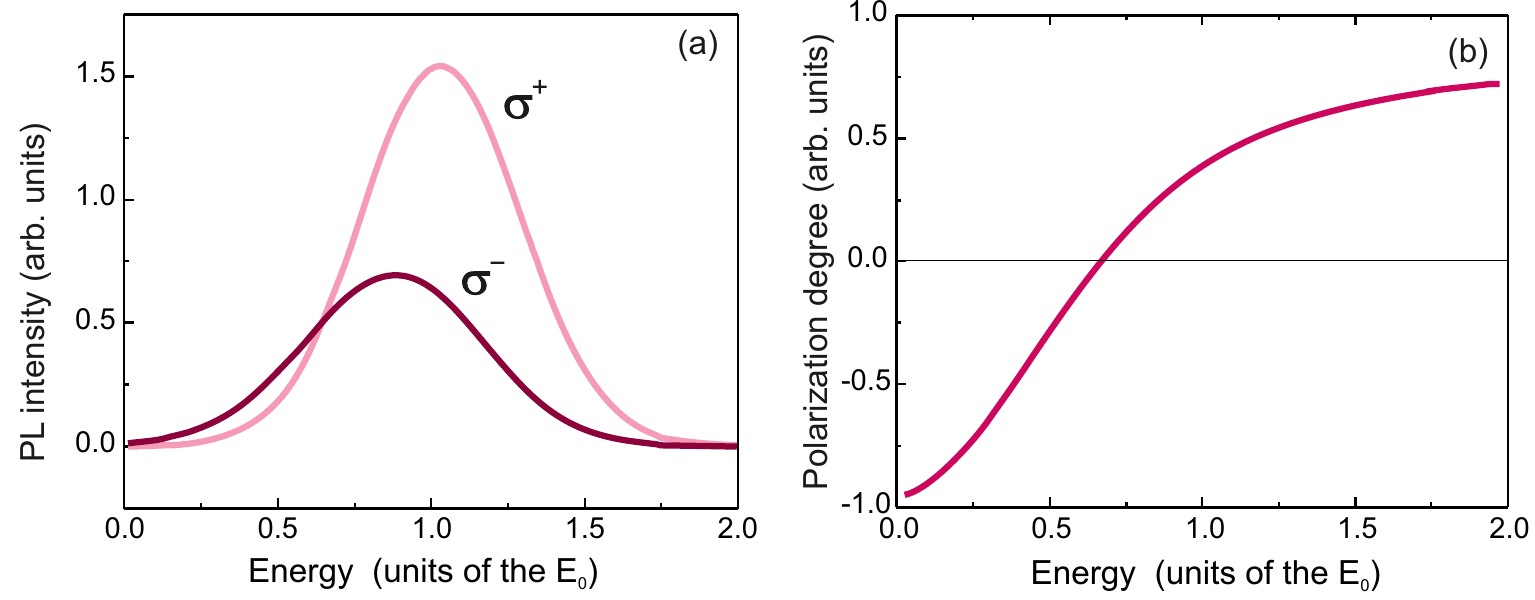}
  \caption{ \textbf{(a)} Emission spectra of an ensemble of quantum dots in two circular polarizations as a function of the exciton quantization energy in the dot (in the units of the central energy   in the Eq.~\eqref{eq7}) under the assumption that the g-factor is described by Eq.~\eqref{eq22},~\eqref{eq23}. \textbf{(b)} Spectral dependence of the degree of polarization of exciton emission. 
}
  \label{fig:5}
\end{figure*}

The obtained dependence of the photoluminescence intensity in two circular polarizations in some fixed magnetic field is presented in Figure~\ref{fig:5}~a. The dependence of the degree of polarization along the contour of the bands is presented in Figure~\ref{fig:5}~b.

We assumed that the parameter $R$ does not depend on the dot size and hence on the exciton energy. If $b=const,$  and   $R$ depends linearly on the energy, the result coincides with the one already obtained. If both parameters $b$  and $R$  depend linearly on energy, then obviously the result is the same. 

\begin{widetext}

\subsection{\label{sec:G}{g-factor}}

Let us find the dependence of the g-factor on the size of the quantum dot. We assume that quantum dots are made on the basis of semiconductors with cubic symmetry. The exciton Hamiltonian in nanostructures based on cubic semiconductors in the effective-mass approximation has the form: 

\begin{equation}
\label{eq13}
    \hat{H}_{exc}= \left( {{\hbar}^2 \hat{\textbf{K}}^2_{e}  \over 2m_e} + V({\textbf{r}_e})\right)\hat{\textbf{I}}  + \left( {{\hbar}^2   \over {2m_0}} (\gamma_1 + {5\over2} \gamma) \hat{\textbf{K}}^2_{h} +V(\textbf{r}_h) - {{e^2}\over { \varepsilon \vert \textbf{r}_e - \textbf{r}_h \vert}}\right)\hat{\textbf{I}}-2 \gamma(\hat{\textbf{J}} \cdot \hat{\textbf{K}}_{h})^2.
\end{equation}
Here: $ \gamma=(2\gamma_2+3\gamma_3)/5 ,\gamma_1  ,\gamma_2  ,\gamma_3 $  are Luttinger parameters, $m_e$ is the effective mass of the electron, $\hat{\textbf{J}}$ is the vector of matrixes of angular momentum 3/2, $\varepsilon$ is the static dielectric permittivity, $\textbf{I}$ is the unit matrix $4 \times 4, V_{e,h}(\textbf{r}_{e,h})$ are the nanostructure potentials for electrons and holes, $\hat{\textbf{K}}_{e,h} = -i \nabla _{e,h}$. 

	We assume that the quantum dot has cylindrical symmetry. Let us separate the motion along $z$ axis and in the plane $(x, y)$ and rewrite Eq.~\eqref{eq13} as: 
 
\begin{multline} 
\label{eq14}
    \hat{H}_{exc}= \left[ {{\hbar}^2   \over 2m_e}\hat{\textbf{K}}^2_{ez} + V_e(z_e)\right]\hat{\textbf{I}}  + \left[ \left( {{\hbar}^2   \over {2m_0}} (\gamma_1 + {5\over2} \gamma) \hat{\textbf{K}}^2_{hz} +V_h(z_h) \right)\hat{\textbf{I}}-\gamma {{\hbar}^2   \over {m_0}}(\hat{\textbf{J}}_z\hat{\textbf{K}}_{hz})^2 \right]+ \\ 
    +\left[ \left( {{\hbar}^2   \over 2m_e}+\hat{\textbf{K}}^2_{e\perp} + V_e(\rho_e)\right)\hat{\textbf{I}}  +  \left( {{\hbar}^2   \over {2m_0}} (\gamma_1 + {5\over2} \gamma) \hat{\textbf{K}}^2_{h\perp} +V_h(\rho_h) - {{e^2}\over { \varepsilon  \sqrt {\rho^2_{ \perp}+{\vert z_h - z_e \vert}^2}}}\right)\hat{\textbf{I}} -\gamma {{\hbar}^2   \over {m_0}}(\hat{\textbf{J}}_\perp \hat{\textbf{K}}_{h\perp})^2\right] - \\
    -\left[ \gamma{{\hbar}^2  \over {m_0}}  \{(\hat{\textbf{J}}_{z} \hat{\textbf{K}}_{hz}) \cdot (\hat{\textbf{J}}_{ \perp} \hat{\textbf{K}}_{h \perp} ) \} \right]. 
\end{multline}
Here: $\textbf{K}^2_\perp = \textbf{K}^2_x + \textbf{K}^2_y $, $\rho^2_\perp =(\overrightarrow{\rho}_e - \overrightarrow{\rho}_h)^2 = {\vert x_h-x_e \vert}^2 + {\vert y_h-y_e \vert}^2 $ , $ \textbf{J}_\perp \textbf{K}_\perp= \textbf{J}_x\textbf{K}_x + \textbf{J}_y\textbf{K}_y $, $\lbrace \textbf{ab} \rbrace = \textbf{ab}+\textbf{ba}$.

Thus, the total Hamiltonian of the exciton in the quantum dot is the sum of fore parts that are in square brackets in Eq.~\eqref{eq14}. Rewire them as: 

  \begin{equation}
\label{eq15}
 \hat{H}=\hat{H}_{1}(z_e)+\hat{H}_{2}(z_h)+\hat{H}_{3}(\rho,\phi,\vert z_e - z_h \vert)+\hat{H}_{4}(\rho,\phi,z_h). 	
\end{equation}				 

For a disc-shaped dots of a small thickness in  $\hat{H}_{3}(\rho,\phi,\vert z_e - z_h \vert)$ one can neglected  $\vert z_e - z_h \vert$  in comparison to $\rho_\perp$  in Eq.~\eqref{eq14}. In this case we can separate the motion of the electron along $z$  axis from its motion in the plane $(x,y)$ . Thus, $\hat{H}_{1}(z_e)$   is not related to all other terms in Eq.~\eqref{eq14}, and we can diagonalize it separately. As a result, we obtain the quantization energy of the electron along the z-axis. It remains to diagonalize the Hamiltonian: 

  \begin{equation}
\label{eq16}
\hat{H}'(\rho,\phi,z_h)=\hat{H}_{2}(z_h)+\hat{H}_{3}(\rho,\phi)+\hat{H}_{4}(\rho,\phi,z_h). 	
\end{equation}				 

The Hamiltonian $\hat{H}_{2}$ describes the quantization of a hole along the  $z$-axis, $\hat{H}_{3}$  describes the internal motion of an electron and a hole in a two-dimensional exciton, and $\hat{H}_{4}$ leads to mixing of the hole motion in the $(x,y)$ plane and along the  $z$-axis. We will consider $\hat{H}_{4}(\rho,\phi,z_h)$  as a perturbation. 

As wave functions of the zero approximation, we choose the eigenfunctions of the Hamiltonian $\hat{H}_{2}(z_h)+\hat{H}_{3}(\rho,\phi)$. They represent by spinors: 

  \begin{equation}
\label{eq17}
\Psi^0_{n,m,M}(\rho, \phi ,z_h)= 
    \begin{pmatrix}
   \phi_M(z_{hh})f_{n,m}(\rho,\phi)\\
  \phi_M(z_{lh})f_{n,m}(\rho,\phi)\\
  \phi_M(z_{lh})f_{n,m}(\rho,\phi)\\
  \phi_M(z_{hh})f_{n,m}(\rho,\phi)
\end{pmatrix}	
\end{equation}				 
Here:  $\phi_M(z_{hh,lh})$  are the eigenfunctions of the Hamiltonian  $\hat{H}_{2}(z_h)$ for the light $(lh)$ holes and heavy $(hh)$ holes along $z$ axis, $f_{n,m}(\rho,\phi)$  are the eigenfunctions of the Hamiltonian of two-dimensional exciton  $\hat{H}_{3}(\rho,\phi)$, $n=0,1,2,...$, $m=0,\pm 1, \pm 2, ..., \pm n$, $M=1,2,...$ are quantum numbers (see additional materials~\ref{appendix}). 

We will consider $\hat{H}_{4}(\rho,\phi,z_h)$  as perturbation. This term gives a nonzero contribution starting only from the second order. The second-order correction to the ground state energy of the 2D heavy hole exciton has the form: 

  \begin{equation}
\label{eq18}
\Delta E_2 =  \sum_{M=1}^{\infty} \sum_{n=1}^{\infty} {{{\vert \langle \Psi^0_{0,0,1}\vert \hat{H}_{4}(\rho,\phi,z_h)    \vert \Psi^0_{n,1,2 M} \rangle \vert}^2} \over {E_0 - E_{n,1,2 M}}}	
\end{equation}				 

Here: $E_0$  is the energy of the ground state of the 2D exciton, $\Psi^0_{0,0,1}(\rho, \phi ,z_h)= f_{0,0} (\rho, \phi)\varphi_1 (z_h)$ is its wavefunction, $E_{n,1,(2 M +1)}$ and $\Psi^0_{n,1,2 M}(\rho, \phi ,z_h)= f_{n,1} (\rho, \phi)\varphi_{2M} (z_h)$ are the energy and wavefunctions of all $p$ states of the 2D exciton on each $2M$ hole level. We assume summation over all $p$-states of exciton and integration over continuum spectrum. 


In the presence of a magnetic field, a replacement is made:$\textbf{K}\Rightarrow \textbf{K}+ {e \over c}\textbf{A}$  . The correction to the energy of the ground state is as follows: 

  \begin{equation}
\label{eq19}
\Delta E_2 =  \gamma {{\hbar}^2 \over m_0} \sum_{M=1}^{\infty} \sum_{n=1}^{\infty} {{{\vert \langle \Psi^0_{0,0,1}\vert \left(  \left( \textbf{K}_x + {e \over c} \textbf{A}_x \right)	\{ J_x J_z \} +  \left( \textbf{K}_y + {e \over c} \textbf{A}_y \right)	\{ J_y J_z \}  \right) \textbf{K}_z \vert \Psi^0_{n,1,2M} \rangle \vert}^2} \over {E_0 - E_{n,1,2M}}} + c.c.	
\end{equation}				 

This correction gives a contribution to the effective mass, to the diamagnetic shift, and to the Zeeman effect. Here we are interested only in the contributions linear in magnetic fields. We consider that the thickness of the QD is small and in Eq.~\eqref{eq19} we can take into account the contribution only from the lower hole quantization level. Then: 

  \begin{equation}
\label{eq20}
 \Delta E_2 \approx 2{\hbar}^2 \left( \gamma \over m_0 \right)  \left( {\gamma \over m_0} {\hbar e \over c}\right) \langle \varphi_1 \vert \textbf{K}_z^2 \vert \varphi_1\rangle \sum_{n=1}^{\infty} {{\langle f_{0,0} \vert \textbf{K}_x\textbf{A}_y + \textbf{A}_x \textbf{K}_y \vert f_{n,1} \rangle} {\langle f_{n,1} \vert \textbf{K}_x\textbf{A}_y + \textbf{A}_x \textbf{K}_y \vert f_{0,0} \rangle} \over {E_0 - E_{n,1}}} \{ J_x J_z \} \{ J_y J_z \} + c.c.	
\end{equation}				 

Approximately this give: 

  \begin{equation}
\label{eq21}
\Delta E_2 \approx {8 \over \pi}  \left( {\gamma \over m_0 } \langle \varphi_1 \vert \textbf{K}_z^2 \vert \varphi_1\rangle \right)  \left( {\gamma \over m_0} {\hbar e \over c}\right) {{\langle f_{0,0} \vert 	\{ \textbf{K}_x\textbf{A}_y + \textbf{A}_x \textbf{K}_y \} \vert f_{0,0} \rangle} \over {E_{(x,y)}+ \Delta (L_Z)} } \left( 7J_z - 4J_z^3\right).	
\end{equation}				 
\end{widetext}

For the wave functions of the 2D exciton $f_{0,0}(\rho,\phi)$  there are two options: 1) the electron and hole Coulomb interaction energy is greater than in-plane $(x,y)$  quantization energy, then the zero approximation wave functions of $\hat{H}_{3}(\rho,\phi)$  are the eigenfunctions of the 2D exciton and the lateral potential of the quantum dot is a small correction, and we will consider it as perturbation; 2) the Coulomb energy is less than the energy of quantization, then the zero approximation eigenfunctions of $\hat{H}_{3}(\rho,\phi)$  are cylinder functions and the Coulomb interaction is considered as a correction. 

In both cases, we obtain that the correction to the Zeeman splitting of the ground state of an exciton with a heavy hole due to "confusion" of the hole motion in the plane $(x,y)$   and perpendicular to the plane has the form:

  \begin{equation}
\label{eq22}
{\Delta E_2 (L_z) }\approx {8 \over {\pi}} (3 \gamma \hbar \omega_c) {E_z \over {E_{(x,y)}+\Delta (LH)}},	
\end{equation}				 
where:  $B$ is magnetic field strength, $\hbar \omega_c=\hbar {{eB}\over{m_e c}}$  electron cyclotron energy,  $E_{(x,y)}$ the sum of 2D exciton binding energy and electron and hole quantization energies in the $(x,y)$ plane in narrow QDs, $\Delta (LH) = E_{lh}-E_{hh}$ ,  $E_{zlh}$ is quantization energy of the light hole along   $z$ axis,   $E_{zhh}$ - is quantization energy of the heavy hole along  $z$ axis, $E_z$  is quantization energy of a particle with mass  $\gamma/m_0$ along  $z$ axis. 

In a magnetic field, the ground state of the exciton experiences the "usual" Zeeman splitting associated with the spin of the electron and the angular momentum of the hole. The obtained correction should be added to this splitting. The total Zeeman splitting of the ground state of the exciton in a quantum dot is equal to: 

  \begin{equation}
\label{eq23}
 \Delta E= \mu g_{e}B+\mu g_{h}^b - \Delta E_2 (L_z) = \mu g_{0}B - \Delta g (E)B.
\end{equation}				 

Here: $g_e+g_h^b=g_0$ is effective  g-factor in the bulk crystal,  $\mu$ is Bohr magneton,  $\Delta g (E)$ is a correction to the g-factor in a quantum dot depending on the exciton energy Eq.~\eqref{eq22}. In cubic crystals $\mu g_h^b \equiv \kappa$ ,  $\kappa$ is Luttinger parameter for Zeeman splitting of heavy holes in a bulk cubic crystal. 

\begin{figure}[h!!]
\centering
 \includegraphics[width=0.95\linewidth]{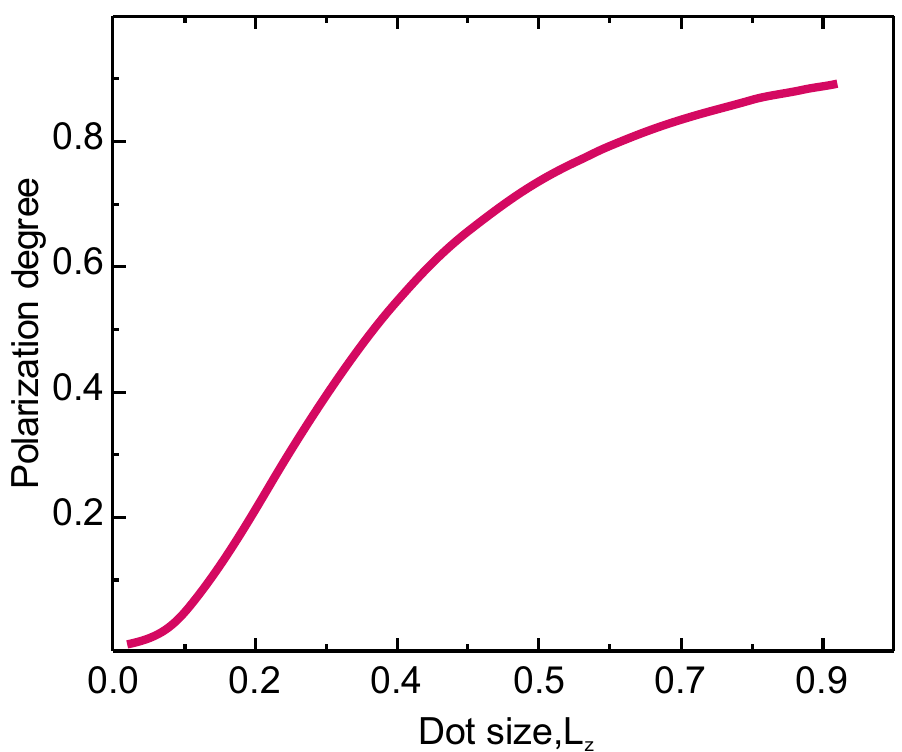}
 \caption{Characteristic, dimensionless dependence of the correction to the exciton g-factor. Here both the g-factor and the quantum well thickness are given in relative units.}
 \label{fig:6}
\end{figure}

The qualitative dependence of the g-factor on $L_z$  in relative units is shown in Figure~\ref{fig:6}. The addition of the g-factor of the bulk exciton leads simply to a shift of zero in this dependence. 

	Thus, for quantum dots of any size, we obtain a universal dependence describing the change in the g-factor as the dot size changes. Similar dependence was obtained for excitons in wide quantum wells~\cite{PhysRevB.81.085208}.

\section{Discussion}

Thus, the change of the exciton g-factor in QD at change of the size of the dot is described by three parameters: (\textit{i})  the quantization energy of heavy holes along  $z$ axis $E_z$ , (\textit{ii})  the difference of quantization energies of light and heavy holes along  $z$ axis $\Delta_{LH}$  and (\textit{iii})  the energy of 2D exciton $E_{(x,y)}$ . All these parameters depend on the QD size. 

A similar result for the g-factor of a free hole in QD was obtained in ~\cite{Efros, 16}. However, in these works the g-factor of a hole not bound into an exciton was considered. 

Generally, the case of interest is the case when $E_z\gg E_{(x,y)},\Delta_{LH}$ . Three limiting cases: 1) $E_z\gg \Delta_{LH}$, 2) $E_z\ll \Delta_{LH}$  and 3)  $E_z\approx \Delta_{LH}$ give the same result as it was shown in paragraph~\ref{sec:1}. The case when  $E_z\ll E_{(x,y)},\Delta_{LH}$ should be considered separately. This case corresponds to an exciton in quantum well wile.

The first case was observed in wide quantum wells for the bulk exciton~\cite{PhysRevB.83.155206}. The second case is characteristic of typical InGaAs/GaAs quantum dots. The third case can be realized in indirect-gap quantum dots~\cite{PhysRevB.78.085323}. 

There can be three cases: 1) the sum of the bulk g-factor $g_0$ and the correction to the g-factor $\Delta g$  in negative, 2) the sum of the bulk g-factor and the correction to the g-factor is positive and 3) total g-factor changes sign depending on the exciton energy. 

If the g-factor does not change sign, the polarization degree  remains of the same sign for all energies. If the g-factor changes sign at some energy, the degree of polarization of photoluminescence also changes sign along the contour of the photoluminescence band. The energy at which the polarization changes sign corresponds to the energy of g-factor zero.

If the g-factor does not depend on the dot size and/or on the radius of the 2D exciton, the splitting of the emission band maxima is small compared to the width of the bands. The bands have the same width. The degree of polarization is constant along the contour of the bands. 

If the g-factor depends on the dot size or on the radius of the 2D exciton, the magnitude of the maxima splitting can be comparable to the width of the bands themselves, like as if the effective g-factor of the exciton has a giant magnitude $g_{eff} \sim 100$, which looks quite unusual for nonmagnetic materials.

This consideration is carried out for equilibrium photoluminescence when thermodynamic equilibrium is established on Zeeman sublevels. The deviation from the equilibrium distribution is taken into account in Eq.~\eqref{eq3} by means of the depolarizing term:

  \begin{equation}
\label{eq24}
	{\tau_0 \over {\tau_0 + \tau_s}}
\end{equation}				 

So, in fact we assume that spin relaxation time  $\tau_s$  is much shorter the lifetime $\tau_0$. In a real situation, there may not be a complete equilibrium. Population ratio in the nonequilibrium case is: 
 
  \begin{equation}
\label{eq27}
	{n_1 \over n_2} = {{\left({e^{-{\Delta E \over {2kT}}}+{\tau_s \over \tau_0}cosh({\Delta E \over{2kT}})}\right)}\over{\left( e^{+{\Delta E \over {2kT}}}+{\tau_s \over \tau_0}cosh({\Delta E \over{2kT}}) \right)}}
\end{equation}

Obviously, (Eq.~\eqref{eq27}) transforms into (Eq.~\eqref{eq1}) if $\tau_s \ll \tau_0$ . However, the consideration of nonequilibrium does not qualitatively change the spectral dependences (Figure~\ref{fig:5}). In the limiting case of the absence of spin relaxation $\tau_s \gg \tau_0$, the difference of the emission spectra in two polarizations disappears. 

In Figure~\ref{fig:4} and ~\ref{fig:5}, attention is attracting to the fact that the lower energy component is less intense than the upper one, which at first glance contradicts the equilibrium distribution of excitons on Zeeman sublevels.

\begin{figure}[t]
\centering
 \includegraphics[width=1\linewidth]{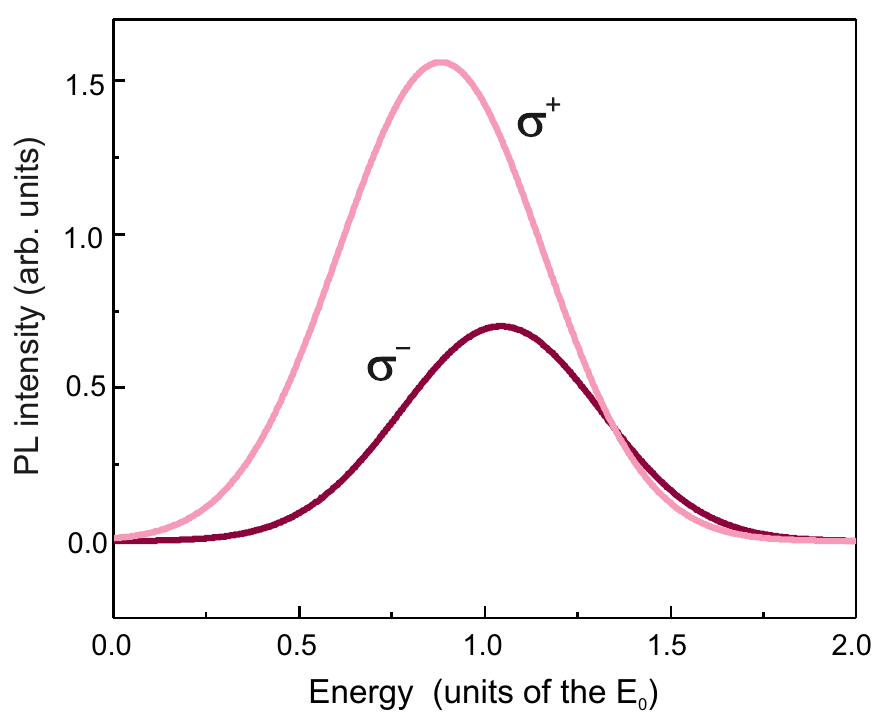}
 \caption{Qualitative view of the emission spectrum of an ensemble of quantum dots in two circular polarizations as a function of the exciton quantization energy in a QD (in the units of the central energy E0 in Eq.~\eqref{eq7}) under the assumption that the g-factor depends linearly on the exciton quantization energy $g_{eff} (E')=g_0 +  \Tilde{g}E'$. }
 \label{fig:7}
\end{figure}

If the g-factor turns to zero at energies below the band maxima ($E^\pm_{max}$  in formula~\eqref{eqn12}), the lower component in the spectrum (Figure~\ref{fig:4} and ~\ref{fig:5}) is less intense; if the g-factor turns to zero at energies above the band maxima, the lower component becomes more intense (Figure~\ref{fig:7}). Parameters of this calculation are: $\mu g_0 B=0.7$, $\mu \Tilde{g} B =0.01(meV)^{-1}.$ 

Thus, the main factor influencing the unusual behavior of magnetically induced polarized luminescence of an ensemble of quantum dots is the dependence of the exciton g-factor on the dispersion of quantum dot sizes.

For the growth of self-organized quantum dots of the InAs/GaAs type, mechanical stresses related to the mismatch of lattice constants are an important factor. Mechanical stresses cause the splitting of light and heavy hole levels $\Delta_{LH}$  depend not only on the thickness but also on the transverse size of the dot and formula~\eqref{eq22} takes all of this into account.

\section{Conclusions}

In this work, the magnetic field induced polarized luminescence of excitons, in an ensemble of quantum dots of different sizes, has been theoretically investigated. It was found that (\textit{i}) the splitting magnitude of photoluminescence bands in a magnetic field in an ensemble of quantum dots can be several orders of magnitude larger than the Zeeman splitting of exciton levels in a single dot, (\textit{ii}) the sign of circular polarization changes along the contour of the bands, (\textit{iii}) the change of the sign of photoluminescence polarization is related to the change of the exciton g-factor sign. A universal formula for the dependence of the g-factor on the quantum dot size is obtained. This dependence is valid both for bulk material and for nanostructures of different types and sizes: quantum wells, quantum filaments and quantum dots. Comparison of the obtained results with experimentally known data is discussed. Changing the sign of the g-factor in an ensemble of quantum dots can strongly influence the effects of mode locking, spin echo and spin coherence in an ensemble of quantum dots~\cite{Greilich}.

\begin{acknowledgments}

L.V.K. thanks Russian Science Foundation (project No. 23-72-10008) for financial support, T.S.Sh. thanks Russian Science Foundation (project No. 22-12-00022). The authors thank Dmitry Smirnov for helpful discussions.
\end{acknowledgments}

\appendix

\begin{widetext}

\section[\appendixname~\thesection]{Additional materials}
\label{appendix}
\renewcommand{\thefigure}{A\arabic{figure}}
\renewcommand{\theequation}{A\arabic{equation}}

We consider only quantum dots with thickness much smaller than the lateral dimension. We assume that the quantum dot has cylindrical symmetry. The exciton Hamiltonian in a quantum dot based on cubic semiconductors in the effective mass approximation has the following form: 

\begin{multline}
\label{eqA1}
\hat{H}_{exc}= \left( {{\hbar}^2   \over 2m_e}\textbf{K}^2_{ez} + V_e(z_e)\right) {\textbf{I}}  + \left( {{\hbar}^2   \over {2m_0}} (\gamma_1 + {5\over2} \gamma) {\textbf{K}}^2_{hz} +V_h(z_h) \right){\textbf{I}}-\gamma {{\hbar}^2   \over {m_0}}({\textbf{J}}_z{\textbf{K}}_{hz})^2 + \\ 
    +\left( {{\hbar}^2 \over m_e} 
 + {\textbf{K}}^2_{e\perp} + V_e(\rho_e) \right) \textbf{I}  +  \left( {{\hbar}^2 \over {2m_0}} (\gamma_1 + {5\over2} \gamma) {\textbf{K}}^2_{h\perp} +V_h(\rho_h)\right){\textbf{I}} - \gamma {{\hbar}^2   \over {m_0}}({\textbf{J}}_\perp \hat{\textbf{K}}_{h\perp})^2 - \\
    - {{e^2}\over { \varepsilon  \sqrt {\rho^2_{ \perp}+{\vert z_h - z_e \vert}^2}}}{\textbf{I}}-\gamma{{\hbar}^2  \over {m_0}}  \{({\textbf{K}}_{hz}{\textbf{J}}_{z} ) ({\textbf{K}}_{h \perp}{\textbf{J}}_{\perp})\}.
\end{multline}				 

Here we have separated the motion along the z-axis and in the plane $(x,y), \gamma=(2\gamma_2+3\gamma_3)/5, \gamma_1, \gamma_2, \gamma_3$ are the Luttinger parameters, $m_e$ is the electron mass, $\overrightarrow{\textbf{J}}=(J_x,J_y,J_z)$  is the vector of matrices of angular momentum 3/2, $\varepsilon$  is a static dielectric permittivity, ${\textbf{I}}$  is the unit matrix $4\times4$, $V_{e,h}(r_{e,h})$ is a quantum dot potentials for electrons and holes, (\textit{we use bold to denote the operators and normal to denote its eigenvalues}). $\textbf{K}^2_\perp = \textbf{K}^2_x + \textbf{K}^2_y $, $\rho^2_\perp =(\overrightarrow{\rho}_e - \overrightarrow{\rho}_h)^2 = {\vert x_h-x_e \vert}^2 + {\vert y_h-y_e \vert}^2 $ , $ \textbf{J}_\perp \textbf{K}_\perp= \textbf{J}_x\textbf{K}_x + \textbf{J}_y\textbf{K}_y ; \textbf{J}_x, \textbf{J}_y, \textbf{J}_z$ are of the following form: 

  \begin{multline}
  \label{eq17}
\textbf{J}_x= 
    \begin{bmatrix}
  0 & \sqrt{3}/2 & 0 & 0 \\
  \sqrt{3}/2 & 0 & 1 & 0 \\
  0 & 1 & 0 & \sqrt{3}/2 \\
  0 & 0 & \sqrt{3}/2 & 0 
  \end{bmatrix},  
  \textbf{J}_y= 
    \begin{bmatrix}
  0 & -i\sqrt{3}/2 & 0 & 0 \\
  i\sqrt{3}/2 & 0 & -i & 0 \\
  0 & i & 0 & -i\sqrt{3}/2 \\
  0 & 0 & i\sqrt{3}/2 & 0 
  \end{bmatrix}, 
  \textbf{J}_z= 
    \begin{bmatrix}
  3/2 & 0 & 0 & 0 \\
  0 & 1/2 & 0 & 0 \\
  0 & 0 & -1/2 & 0 \\
  0 & 0 & 0 & -3/2 
  \end{bmatrix}
  \nonumber
\end{multline}				 

The first summand in round brackets describes the motion of the electron along the axis $z$ , the second and third summands in round brackets describe the motion of the holes along $z$ , similarly, the three summands in the second line describe the motion of the electron and the hole in the plane $(x,y)$, the first summand in the third line describes the Coulomb interaction between the electron and the hole. The last summand in Eq.~\eqref{eqA1} describes the mixing of the hole motion in the plane $(x,y)$ and along the axis $z$. This summand corresponds to the non-diagonal elements of the matrix of the Hamiltonian~\eqref{eqA1}. The matrix of the Hamiltonian can be partially diagonalized and consider this summand as a perturbation. 

We will assume that the thickness of the dot is much smaller than its diameter and Bohr radius of the exciton. Usually, in self-organized quantum dots it is so. Then in the Coulomb interaction energy of the electron and hole the summand  ${\vert z_h - z_e \vert}^2$ can be neglected compared to $\rho^2$ . In this case the motion of the electron along the axis is in no way related to all others and one can consider the quantization of electrons and holes along the $z$-axis independently of each other and of their motion in the plane $(x,y)$. 

The quantization of electron motion along the axis is described by Eq: 
\begin{equation}
\label{eqA2}
\left[ {{\hbar}^2   \over 2m_e}\textbf{K}^2_{ez} + V_e(z_e)-E_{Ne} \right]\varphi_N(z_e)=0.	
 \end{equation} 

If the potential $V_e(z_e)$  for an electron is infinitely deep, the energy levels and wave functions are of the form: 
\begin{equation}
\label{eqA3}
E_{Ne} = {{{\pi}^2{\hbar}^2N^2} \over 2m_e L^2 } ;
\varphi_N (z_e) = \sqrt{2 \over L} \cos \left(N{\pi  \over L}z_e\right).
\end{equation} 

The quantization of the hole motion along the axis $z$ is described by Eq: 
\begin{equation}
\label{eqA4}
\left[ {{\hbar}^2   \over {2m_0}} (\gamma_1 + {5\over2} \gamma) {\textbf{K}}^2_{hz} -\gamma {{\hbar}^2   \over {m_0}}({\textbf{J}}_z{\textbf{K}}_{hz})^2 +V_h(z_h)-E_{Mh}\right]\varphi_M(z_h)=0.
\end{equation} 

For an infinitely deep potential $V_h(z_h)$  we obtain: 
\begin{equation}
\label{eqA5}
E_{Mh} = {{{\pi}^2{\hbar}^2M^2} \over 2m_{hh,lh} L^2 } ;
\varphi_M (z_{hh,lh}) = \sqrt{2 \over L} \cos \left(M{\pi  \over L}z_{hh,lh}\right).
\end{equation} 

Here:  $m_{hh}={m_0\over{\gamma_1-2\gamma}}$ ,  $m_{lh}={m_0\over{\gamma_1+2\gamma}}$ . 

For the exciton we will neglect non-diagonal elements in the hole dispersion, then for 2D exciton in the plane $(x,y)$ we obtain the equation: 
\begin{equation}
\label{eqA6}
\left[ \left( {{\hbar}^2   \over {2m_e}}+{\textbf{K}}_{e\perp}^2 +V_e(\rho_e)\right)+ \left( {{\hbar}^2   \over {2m_0}} (\gamma_1 \pm 2\gamma){\textbf{K}}_{h\perp}^2+V_h(\rho_h)-{e^2\over {\varepsilon \rho}}\right) -E_\rho\right]f(\rho_e,\rho_h)=0.
\end{equation} 

We consider the motion of the exciton center-of-mass in the plane as a small correction to the energy. In polar coordinates for the 2D exciton we obtain Eq: 
\begin{equation}
\label{eqA7}
{{\hbar}^2 \over {2\mu_\pm} }\left({1 \over \rho } {\partial\over {\partial\rho}}\rho {\partial\over {\partial\rho}} + {1 \over \rho^2}{\partial^2\over {\partial{\phi}^2}}
 - {{e^2} \over {\varepsilon \rho}}+E\right)f(\rho, \varphi)=0.
\end{equation} 

Here: ${1 \over \mu_\pm} = {1 \over m_e }+{{\gamma_1 \pm 2\gamma} \over m_0}$   is reduced mass of heavy and light excitons, $\phi=\arctan \left({{y_e-y_h}\over {x_e-x_h}}\right)$, $\rho = \sqrt{{\vert{x_e-x_h}\vert}^2+{\vert{y_e-y_h}\vert}^2}.$
Solving equation~\eqref{eqA7}, we find the energy levels: 
\begin{equation}
\label{eqA8}
E_{n,m}=-{{\mu e^4}\over {2{\hbar}^2 \varepsilon^2}}{1\over{(\vert m \vert + 1/2 + n_r)^2}}=-{e^2 \over {2 \varepsilon a_B}} {1 \over{(n+1/2)^2}}
\end{equation} 
and wave functions~\cite{10.1119/1.1971931}.

\begin{equation}
\label{eqA9}
f_{n,m}(\rho, \varphi)={{e^{im\varphi}\over {\sqrt{\pi}}} {2 \over a_B}{1 \over{2n+1}}} \sqrt{2\over{2n+1}} \sqrt{{(n- \vert m \vert)!} \over {(n+ \vert m \vert)!}} (2\xi)^{\vert m \vert} e^{-\xi} L_{n-\vert m \vert}^{2\vert m \vert} (\xi);
\xi = {2 \over{2n+1}}{\rho \over a_B}.
\end{equation} 

The summand mixes hole motion in the $(x,y)$ plane and motion perpendicular to the plane:
\begin{equation}
\label{eqA10}
\gamma {\hbar^2 \over m_0} \left[ \textbf{K}_x \{ {\textbf{J}}_{x}{\textbf{J}}_{z}\}+ \textbf{K}_y \{ {\textbf{J}}_{y} {\textbf{J}}_{z}\}\right]\textbf{K}_z
\end{equation} 

In the magnetic field a generalized momentum is introduced $\textbf{K}\Rightarrow \textbf{K}+ {e \over c}\textbf{A}$  , then~\eqref{eqA10} is written: 
\begin{equation}
\label{eqA11}
\gamma {\hbar^2 \over m_0} \left[ \left(\textbf{K}_x + {e \over c}\textbf{A}_x \right)\{ {\textbf{J}}_{x}{\textbf{J}}_{z}\}+ \left( \textbf{K}_y + {e \over c}\textbf{A}_y \right)\{ {\textbf{J}}_{y} {\textbf{J}}_{z}\}\right]\textbf{K}_z
\end{equation} 

This summand will be considered by perturbation theory. It gives a contribution only in the second order. For the zero approximation we have two options. 

I) \textit{Large diameter dots and wetting layer}. 

For such dots the quantization energy along the $z$-axis is greater than all other energies. We neglect the influence of the boundaries of a point in the plane $(x,y)$. It is in essence a case of a quantum well. That is: $\rho^2 \gg \vert z_h-z_e \vert^2, V_e(\rho_e)+V_h(\rho_h)=0$.

As a zero approximation, we choose wave functions of 2D exciton and quantization of a heavy hole along the $z$-axis. In the exciton we neglect the difference of light and heavy holes. We consider the summand~\eqref{eqA11} as a perturbation. Since the perturbation refers only to holes, the quantization of the electron along the $z$-axis does not affect anything. Wave functions of the zero approximation are: 
\begin{equation}
\label{eqA12}
\Psi^0_{n,m,M}(\rho, \phi ,z_{hh})=\varphi_M(z_h)f_{n,m} (\rho,\varphi).
\end{equation} 

The correction to the ground state energy associated with the perturbation~\eqref{eqA11} has the form: 
\begin{equation}
\label{eqA13}
\Delta E_2 =  \gamma {{\hbar}^2 \over m_0} \sum_{M=1}^{\infty} \sum_{n=1}^{\infty} {{{\vert \langle \Psi^0_{0,0,1}\vert \left(  \left( \textbf{K}_x + {e \over c} \textbf{A}_x \right)	\{ J_x J_z \} +  \left( \textbf{K}_y + {e \over c} \textbf{A}_y \right)	\{ J_y J_z \}  \right) \textbf{K}_z \vert \Psi^0_{n,1,2M} \rangle \vert}^2} \over {E_1 - E_{n,1,2M}}} + c.c.	
\end{equation} 

We consider that the thickness of the QD is small and in~\eqref{eqA13} we can take into account the contribution only from the lower hole quantization level. Then the linear part of the magnetic field in this correction has the form: 

  \begin{equation}
 \Delta E_2 \approx 2{\hbar}^2 \left( \gamma \over m_0 \right)  \left( {\gamma \over m_0} {\hbar e \over c}\right) \langle \varphi_1 \vert \textbf{K}_z^2 \vert \varphi_1\rangle \sum_{n=1}^{\infty} {{\langle f_{0,0} \vert \textbf{K}_x\textbf{A}_y + \textbf{A}_x \textbf{K}_y \vert f_{n,1} \rangle} {\langle f_{n,1} \vert \textbf{K}_x\textbf{A}_y + \textbf{A}_x \textbf{K}_y \vert f_{0,0} \rangle} \over {E_0 - E_{n,1}}} \{ J_x J_z \} \{ J_y J_z \} + ...
 \nonumber
\end{equation}				 

Approximately this give: 
\begin{equation}
\label{eqA14}
\Delta E_2 \approx {8 \over \pi}  \left( {\gamma \over m_0 } \langle \varphi_1 \vert \textbf{K}_z^2 \vert \varphi_1\rangle \right)  \left( {\gamma \over m_0} {\hbar e \over c}\right) {{\langle f_{0,0} \vert 	\{ \textbf{K}_x\textbf{A}_y + \textbf{A}_x \textbf{K}_y \} \vert f_{0,0} \rangle} \over {E_{(x,y)}+ \Delta_{LH}} } \left( 7J_z - 4J_z^3\right).
\end{equation} 

For quantum dots of large diameter, when the diameter of the dot is much larger than the Bohr radius 2D of the exciton, we can consider $E_{(x,y)}=Ry^{2D}$. Then the correction to the Zeeman splitting of holes related to their quantization will be:

\begin{equation}
\label{eqA15}
{\Delta E^\perp_2 (L_z,a_B) }\approx {8 \over {\pi}} (3 \gamma \hbar \omega_c) {E_z \over {Ry^{2D}+\Delta (L_z)}},
\end{equation} 
here: $\Delta (L_z)=E_{zlh}-E_{zhh}, E_{zlh}  $ is quantization energy of a light hole along the axis   $z$, $E_{zhh}$   is quantization energy of a light hole along the  axis $z$, $E_z$  is quantization energy of a hole with mass  $\gamma / m_0$ along the axis $z$  ,  $\hbar \omega_c$ is cyclotron energy of a free electron,  $Ry$ is exciton Rydberg $Ry^{2D} = {{2\mu e^4}\over{\hbar^2 \varepsilon^2}}$ . 

II)  \textit{Small diameter points}. 

We consider that the quantization energy along the $z$-axis is larger than all other energies. We will consider the Coulomb interaction ${{e^2}\over { \varepsilon  \sqrt {\rho^2+{\vert z_h - z_e \vert}^2}}}$   of the electron and hole as a correction to the energy of the ground state. Then we obtain the independent motion of the electron and hole in the cylindrical dot.
\begin{equation}
\label{eqA16}
\left[ \left( {{\hbar}^2   \over {2m_e}}+{\textbf{K}}_{e\perp}^2 +V_e(\rho_e)\right)+ \left( {{\hbar}^2   \over {2m_0}} (\gamma_1 \pm 2\gamma){\textbf{K}}_{h\perp}^2+V_h(\rho_h)-{e^2\over {\varepsilon \rho}}\right) -E_\rho\right]f(\rho_e,\rho_h)=0.
\end{equation} 

In polar coordinates we obtain:
\begin{equation}
\label{eqA17}
{{\hbar}^2 \over {2m_{e,h}} }\left({1 \over \rho } {\partial\over {\partial\rho}}\rho {\partial\over {\partial\rho}} + {1 \over \rho^2}{\partial^2\over {\partial{\phi}^2}}
 +V_{e,h}(\rho)+E\right)f(\rho, \varphi)=0.
\end{equation} 

Let the radius of a point be equal to $R_0$ . The wave functions in the plane $(x,y)$  are Bessel functions:
\begin{equation}
\label{eqA18}
f_{n,m}(\rho,\varphi)=\sqrt{1 \over{\pi R^2_0}}{{J_m(\xi_{n,m}\cdot\rho/2R_0)}\over{J_{m+1}(\xi_{n,m}})}e^{im\varphi}
\end{equation} 

Energy Levels: 
\begin{equation}
\label{eqA19}
E^\perp_{(n,m)}={{\hbar}^2 \over {2m_{e,h}} }{\xi^2_{n,m}\over R^2_0}.
\end{equation} 

Here $ \xi_{n,m} $ are the roots of the Bessel functions. The correction to the Zeeman splitting in this case has a form similar to~\eqref{eqA15}: 
\begin{equation}
\label{eqA20}
{\Delta E_2 (L_z, R_0) }\approx {8 \over {\pi}} (3 \gamma \hbar \omega_c) {E_z \over {E^\perp_{(0,0)}+\Delta (L_z)}}.
\end{equation} 

We see that in both cases the total Zeeman splitting of the ground state of the exciton in the quantum dot is equal: 
\begin{equation}
\label{eqA21}
\Delta E= \mu g_{e}B+\mu g_{h}^b B - \Delta E_2 (L_z),
\end{equation} 
where: $g_e$  is electron  $g$-factor, $g_h^b$  is hole  $g$-factor in bulk InAs, $\mu$  is Bohr magneton. 

Thus, for quantum dots of any size, we obtain a universal dependence describing the change of the $g$-factor at the change of the dot size. The same dependence was obtained for excitons in wide quantum wells~\cite{https://doi.org/10.1002/pssc.201000847} and for holes  in quantum dots and quantum wires, neglecting their Coulomb interaction with electrons~\cite{16,Efros}. 
\end{widetext}
\nocite{*}

\bibliography{apssamp}

\end{document}